\documentclass[aps,prl,preprint,superscriptaddress]{revtex4-1}

\usepackage{epsfig}
\usepackage{graphics}
\usepackage{amsfonts}
\usepackage{mathrsfs}
\usepackage{amsmath}% needed for subequations
\usepackage{color}
\usepackage{natbib}

\usepackage{textcomp}
\usepackage{graphicx}
\usepackage{bm}% bold maths
\usepackage{amssymb}
\usepackage{ulem}
\usepackage{xspace}
\usepackage{epstopdf}
\usepackage{dcolumn}% Align table columns on decimal point
\usepackage{subfigure}
\usepackage{multirow}
\usepackage{diagbox}
\usepackage{array}
\usepackage{geometry}

 \usepackage{makecell}
\usepackage[colorlinks=true, letterpaper=true, pdfstartview=FitV, linkcolor=blue, citecolor=blue, urlcolor=blue]{hyperref}

\begin{document}

\title{\textbf{Observation of superconductivity in a nontrivial $\mathcal{Z}_2$ approximant quasicrystal}}

\author{Pavan Kumar Meena $^*$\footnote[0]{*These authors contributed equally to this work.}}
\affiliation{Department of Physics, Indian Institute of Science Education and Research Bhopal, Bhopal, 462066, India}

\author{Rahul Verma $^*$\footnote[0]}
\affiliation{Department of Condensed Matter Physics and Materials Science, Tata Institute of Fundamental Research, Mumbai 400005, India}

\author{Arushi}
\affiliation{Department of Physics, Indian Institute of Science Education and Research Bhopal, Bhopal, 462066, India}

\author{Sonika Jangid}
\affiliation{Department of Physics, Indian Institute of Science Education and Research Bhopal, Bhopal, 462066, India}

\author{Roshan Kumar Kushwaha}
\affiliation{Department of Physics, Indian Institute of Science Education and Research Bhopal, Bhopal, 462066, India}

\author{Rhea Stewart}
\affiliation{ISIS Facility, STFC Rutherford Appleton Laboratory, Harwell Science and Innovation Campus, Oxfordshire, OX11 0QX, UK}

\author{{Adrian D. Hillier}}
\affiliation{ISIS Facility, STFC Rutherford Appleton Laboratory, Harwell Science and Innovation Campus, Oxfordshire, OX11 0QX, UK}

\author{Bahadur Singh $^{\dag}$}

\affiliation{Department of Condensed Matter Physics and Materials Science, Tata Institute of Fundamental Research, Mumbai 400005, India}

\author{{Ravi Prakash Singh $^{\dag}$ \footnote[0]{$^{\dag}$Corresponding emails: bahadur.singh@tifr.res.in, rpsingh@iiserb.ac.in}}} 
\affiliation{Department of Physics, Indian Institute of Science Education and Research Bhopal, Bhopal, 462066, India}

\maketitle

\clearpage
\newpage

\section*{Abstract}
\textbf{Superconductivity and nontrivial topology are highly sought-after phenomena in quantum materials. While many topological crystalline materials have been found to exhibit superconductivity, their presence in quasicrystals - materials with a unique aperiodic yet ordered structure - has remained largely unexplored. In this work, we report the discovery of superconductivity in a monoclinic approximant to the decagonal quasicrystal Al$_{13}$Os$_{4}$, that exhibits a high superconducting transition temperature and a nontrivial electronic structure. The resistivity, magnetization, specific heat, and $\mu$SR measurements confirm superconductivity with a critical temperature of $\sim5.47$ K. Detailed electronic structure and symmetry analysis reveal nontrivial state with $\mathcal{Z}_{2}=1$ and spin-polarized conducting surface states. Importantly, we identify three-dimensional saddle point van Hove singularities with substantial flat energy dispersion at the Fermi level, which can enhance superconductivity. Our results highlight a rich interplay between superconductivity and nontrivial electronic states in Al$_{13}$Os$_{4}$, demonstrating it as a unique platform for exploring unconventional superconducting states in quasicrystalline materials.}\\

\section*{Introduction}
Quasicrystals (QCs), first discovered by Shechtman {\it et al.} in 1984, exhibit long-range atomic order without periodicity, thereby occupying a distinct structural regime between periodic and amorphous materials. This long-range order manifests in sharp Bragg peaks resulting from crystallographically forbidden rotational symmetries, such as 5-, 8-, 10-, and 12-fold symmetries. Initially identified in intermetallic alloys, QCs have since been observed in a variety of systems, though they remain less explored compared to periodic and amorphous materials~\cite{shechtman1984metallic}. QCs exhibit intriguing electronic and magnetic properties, including long-range magnetic order, low-temperature order-disorder transitions, pseudogaps near the Fermi level, and unconventional magnetic behavior~\cite{deguchi2012quantum, hafner1992electronic, nayak2012bulk, tamura2021experimental}. 

Advances in theoretical explorations and experimental characterization have recently revealed diverse phenomena in QCs, such as non-trivial bulk and surface states, Fulde-Ferrell-Larkin-Ovchinnikov (FFLO) superconductivity, nematic order, and localized Cooper pair distributions~\cite{chen2020higher, fan2022topological, ghadimi2021topological, sakai2019exotic, araujo2019conventional, liu2024nematic}. However, these phenomena challenge conventional solid-state theories, such as Bloch's theorem and topological band theory, which rely on the translational symmetry, and the Bardeen-Cooper-Schrieffer (BCS) theory of superconductivity, which assumes Cooper pair formation with zero center-of-mass momentum~\cite{sakai2017superconductivity}. QCs are often accompanied by a periodic phase resulting from their projection along rational directions, known as approximants. These approximants, while periodic with ordered motifs and symmetries, share local similarities with QCs and serve as a bridge between periodic and aperiodic structures. This periodic framework helps simplify the understanding of the complex properties of QCs~\cite{graebner1987specific, wagner1988specific}. For instance, superconductivity has been observed in QCs, such as a Bergmann-type Al-Mn-Zn alloy with a transition temperature ($T_C$) around 0.05 K~\cite{kamiya2018discovery}, and a van der Waals-layered two-dimensional (2D) dodecagonal QC with a $T_C$ of 1 K~\cite{tokumoto2024superconductivity, terashima2024anomalous} as well as in their approximants. Such studies infer that superconductivity remains robust under variations of the atomic arrangement from periodic to quasiperiodic, though the superconducting $T_C$  in QCs remains relatively low.

Meanwhile, various theoretical studies have predicted topological phases and their candidate materials in QCs by exploiting the periodic symmetries of approximants or considering long-range quasicrystalline order as arising from periodic structures in higher-dimensional spaces~\cite{Kraus_TQC1, Kraus_TQC2, cain2020layer}. Despite these predictions, no quasicrystalline materials have yet shown both high-temperature superconductivity and a nontrivial electronic structure. Notably, decagonal QCs and their approximants exhibit quasiperiodic ordering in the plane and periodicity along one axis. They can provide a particularly unique system for exploring superconductivity and nontrivial topological phenomena in materials that combine quasiperiodicity and periodicity \cite{singh2023topology}. However, the superconductivity or nontrivial topological structure in decagonal QCs remains elusive. 

In this work, we report the discovery of superconductivity in the decagonal approximant Al$_{13}$Os$_{4}$~\cite{zhang1995al13, li1995structure}, and demonstrate its nontrivial electronic structure. Through comprehensive experimental investigations, including magnetization, resistivity, specific heat, and advanced $\mu$SR measurements, we observe bulk superconductivity with a $T_C$ of 5.47(2) K and preserved time reversal symmetry in superconducting ground state. Despite its complex structure, the superconducting properties exhibit good agreement with the weak coupling limit of the BCS theory. Our detailed bulk and surface state calculations reveal nontrivial $Z_2$ invariants and spin-polarized topological surface states that cross the Fermi level. We also identify 2D Fermi sheets and saddle-point van Hove singularities at the Fermi level, which may enhance superconducting $T_C$ in Al$_{13}$Os$_{4}$. The combination of high $T_C$, bulk superconductivity, and nontrivial topological features makes Al$_{13}$Os$_{4}$ as a promising candidate for exploring topological superconductivity. Our work opens new avenues for studying QCs and their approximants as materials for unconventional superconductivity and topological phenomena in quantum condensed matter physics. 

\section*{Results}

\subsection*{Crystal structure}

The crystal structure of the approximant decagonal QCs Al$_{13}$Os$_{4}$ adopts a low-symmetry monoclinic structure with space group $C2/m$ (No. 12) and $C_{2h}$ point-group symmetry (see Figs.~\ref{Fig1}(a)-(c)),isostructural to the $Y$ phase of Al$_{13-x}$(Co$_{4(1-y)}$Ni$_{4y}$) or Al$_{11}$Co$_{4}$~\cite{zhang1995al13, li1995structure}. Fig.~\ref{Fig1}(d) shows the powder X-ray diffraction pattern (XRD) and Rietveld refinement (detailed structural parameters and Wyckoff positions given in the SM). 

The structure comprises of two quasi-periodic layers stacking along the $\vec{b}$ direction with a periodicity of $\sim$ 4 $\text{\AA}$. A quasiperiodic atomic layer in the $\vec{b}=0$ plane connects to another layer at $\vec{b} = 1/2$ plane through a shift in the lattice constant by $a/2$ to form a periodic unit cell (Fig.~\ref{Fig1}(b)). It consists of slightly distorted pentagons and rhombi (Fig.~\ref{Fig1}(a)), with an Al atom occupying one vertex of the pentagon and Os atoms occupying the remaining vertices. Three Al atoms also lie in this Os-Al pentagon. Smaller Al pentagons, made up of five Al atoms with an Os atom at the center, also appear in the adjacent layers. Despite slight differences in Al and Os occupancy, the local structural configuration in the two quasi-periodic layers is nearly identical. In the periodic approximant phase, pentagonal elements are distorted, unlike the regular pentagons in the decagonal quasicrystal phase, representing the primary difference between the two phases~\cite{zhang1995al13, steurer1993structure}. Figure~\ref{Fig1}(c) shows the repetitive arrangement of distorted pentagons and rhombi that tile in the plane to generate the periodic crystal structure. It is important to note that a decagonal QC with two quasiperiodic layers with a periodicity length of about 0.4 $nm$ is the most anisotropic case, whereas other decagonal QCs have four, six, and eight layers in a periodic unit with periodicities of about 0.8, 1.2, and 1.6 $nm$, respectively, exhibiting progressively lower anisotropies.

\subsection*{Bulk superconductivity}

\begin{figure*}
\includegraphics[width=0.98\textwidth]{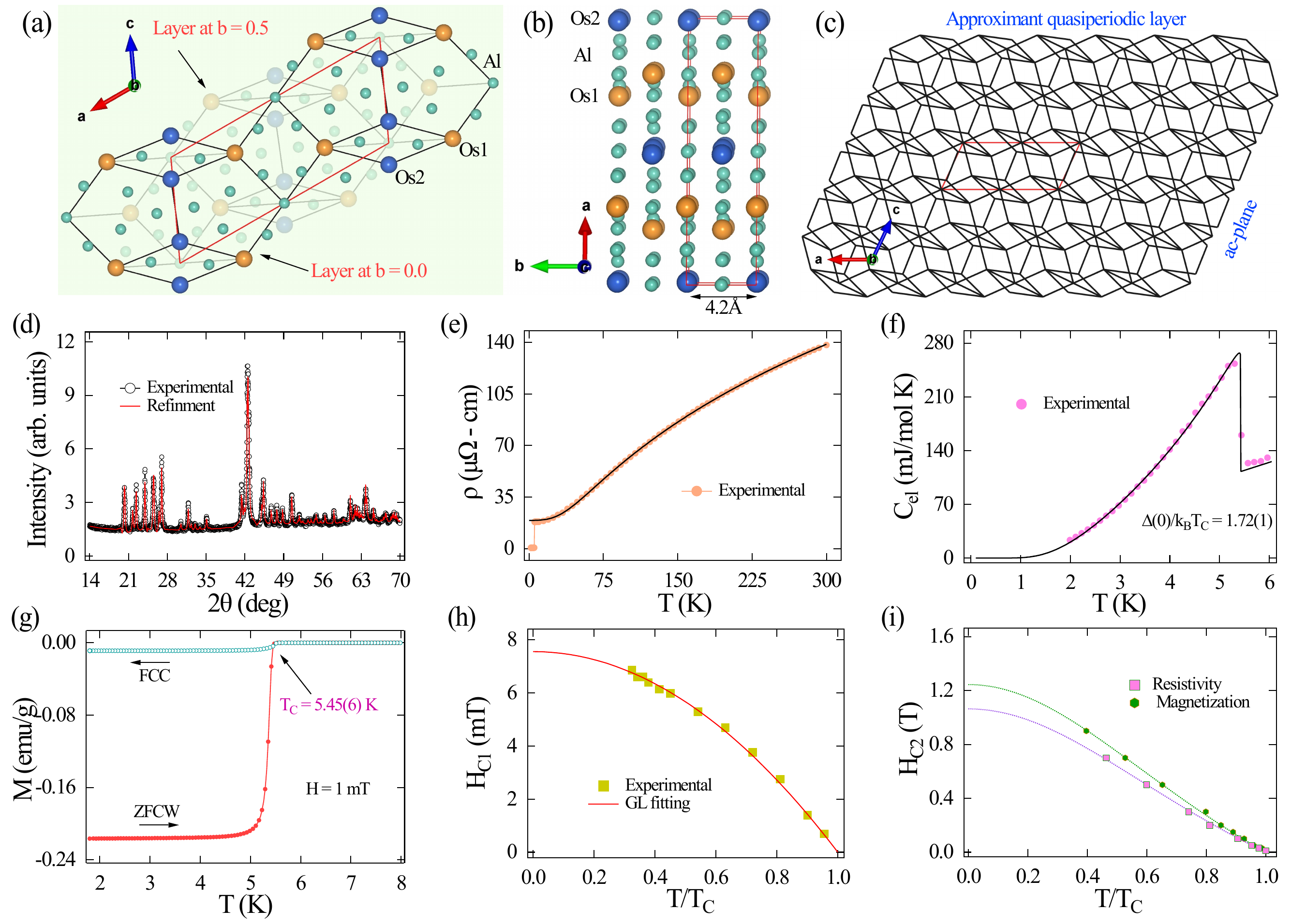}
\caption {\textbf{Crystal structure and bulk superconductivity characterization of the decagonal quasicrystal approximant Al$_{13}$Os$_{4}$.} (a) Crystal structure of the decagonal approximant characterized with distorted pentagons and rhombi. The aperiodic layer at $\vec{b} = 0.0$ connects to the layer at $\vec{b} = 1/2$ to form an approximant decagonal structure. The red rhombus outlines the unit cell. (b) Periodic repetition of the unit cell along the $\vec{b}$ direction, with a periodicity of 4.2 \AA. (c) Two-dimensional quasi-periodic lattice in the $ac$-plane with a unit cell highlighted in red color. (d) Powder X-ray diffraction pattern (black markers) and Rietveld refinement (red line) of Al$_{13}$Os$_{4}$. (e) Temperature dependence of resistivity $\rho(T)$ at $H$ = 0 T, showing the superconducting transition at $T_C$ = 5.47(2) K. (f) Temperature-dependent electronic specific heat ($C_{el}$), with the fitting of data using weak-coupling BCS model (solid black line). (g) Temperature-dependent ZFCW and FCC magnetization at an external field of 1.0 mT, demonstrating clear superconducting transition. (h) and (i) Temperature-dependent lower $H_{C1}$ and upper $H_{C2}$ critical fields, revealing the large upper critical field.}
\label{Fig1} 
\end{figure*}

Figure~\ref{Fig1}(e) presents the temperature-dependent electrical resistivity $\rho(T)$ of Al$_{13}$Os$_4$ under zero-applied magnetic field. A sharp decrease in $\rho(T)$ to zero at a transition temperature $T_C = 5.47(2)$ K confirms the onset of superconductivity, the highest T$_C$ observed to date in any known quasicrystals and approximants (see SM for details). The high-temperature behavior of $\rho(T)$ is metallic, which is well described by Wiesmann's parallel resistor model (black solid line in Fig.\ref{Fig1}(e)). The residual resistivity ratio, $\rho({300})/\rho({10}) = 7.5$, further corroborates the metallic nature of the system (additional details on fitting parameters, Kadowski-Woods ratio, and the Hall measurement data are given in SM).

Specific heat measurements reveal a lambda-shaped anomaly observed at $T_{C}$ = 5.44(2)K, confirming bulk superconductivity consistent with resistivity and magnetization data (see SM). Fig.~\ref{Fig1}(f), shows temperature-dependent electronic specific heat, fitted with the BCS model. The fitting results yield a superconducting gap value ($\frac{\Delta(0)}{k_{B}T_{C}}$) of 1.72(1), indicating weak coupling BCS superconductivity in Al$_{13}$Os$_4$. Additionally, the density of states at the Fermi level $D({E_{F}})$ estimated to be 9.15(4) states eV$^{-1}$ f.u.$^{-1}$, and the Debye temperature $\theta_{D}$, 340(4) K, comparable to those found in elemental Os and Al, the approximant phase of quasicrystals, and other complex intermetallic compounds~\cite{graebner1987specific}. Using the McMillan model, the electron-phonon coupling strength $\lambda_{e-ph}$ is estimated to be 0.63(1), indicating weakly coupled superconductivity in Al$_{13}$Os$_4$. A detailed mathematical description of this analysis is provided in the SM.

Figure~\ref{Fig1}(g) displays the temperature-dependent magnetization of Al$_{13}$Os$_4$ measured in a 1 mT applied field. Both the zero field-cooled warming (ZFCW) and the field-cooled cooling (FCC) curves exhibit a sharp diamagnetic response below $T_C$ = 5.45(6) K, confirming the Meissner effect and the onset of superconductivity. The observed difference between the ZFCW and FCC curves is attributed to flux pinning, a characteristic of type-II superconductors. The AC susceptibility and magnetization loop further confirm the superconducting transition and vortex pinning (see SM). Using $M$ versus $H$ and $M$ versus $T$ data and applying the Ginzburg-Landau (GL) relations (Figs.~\ref{Fig1}(h) and \ref{Fig1}(i)) the lower and upper critical fields are determined to be obtained as $H_{C1}$ = 7.5(4) mT and $H_{C2}$ = 1.24 (1) T, respectively. From these values, two important length scales are extracted, the penetration depth $\lambda_{GL}$(0) = 249(5) nm and coherence length $\xi_{GL}$(0) = 16.2(9) nm. The GL parameter $\kappa_{GL} = \lambda_{GL}(0)/\xi_{GL}(0)$ = 15.3(1) indicates strong type-II superconductivity. The thermodynamic critical field is obtained as $H_{C}$ = 58.3(8) mT. Moreover, the calculated Maki parameter $\alpha_{M}$ = 0.14 suggests that the Pauli limiting effect has a negligible influence on superconductivity, with orbital limiting being the dominant factor.

\subsection*{$\mu$SR results}

\begin{figure*}
\includegraphics[width=0.98\textwidth]{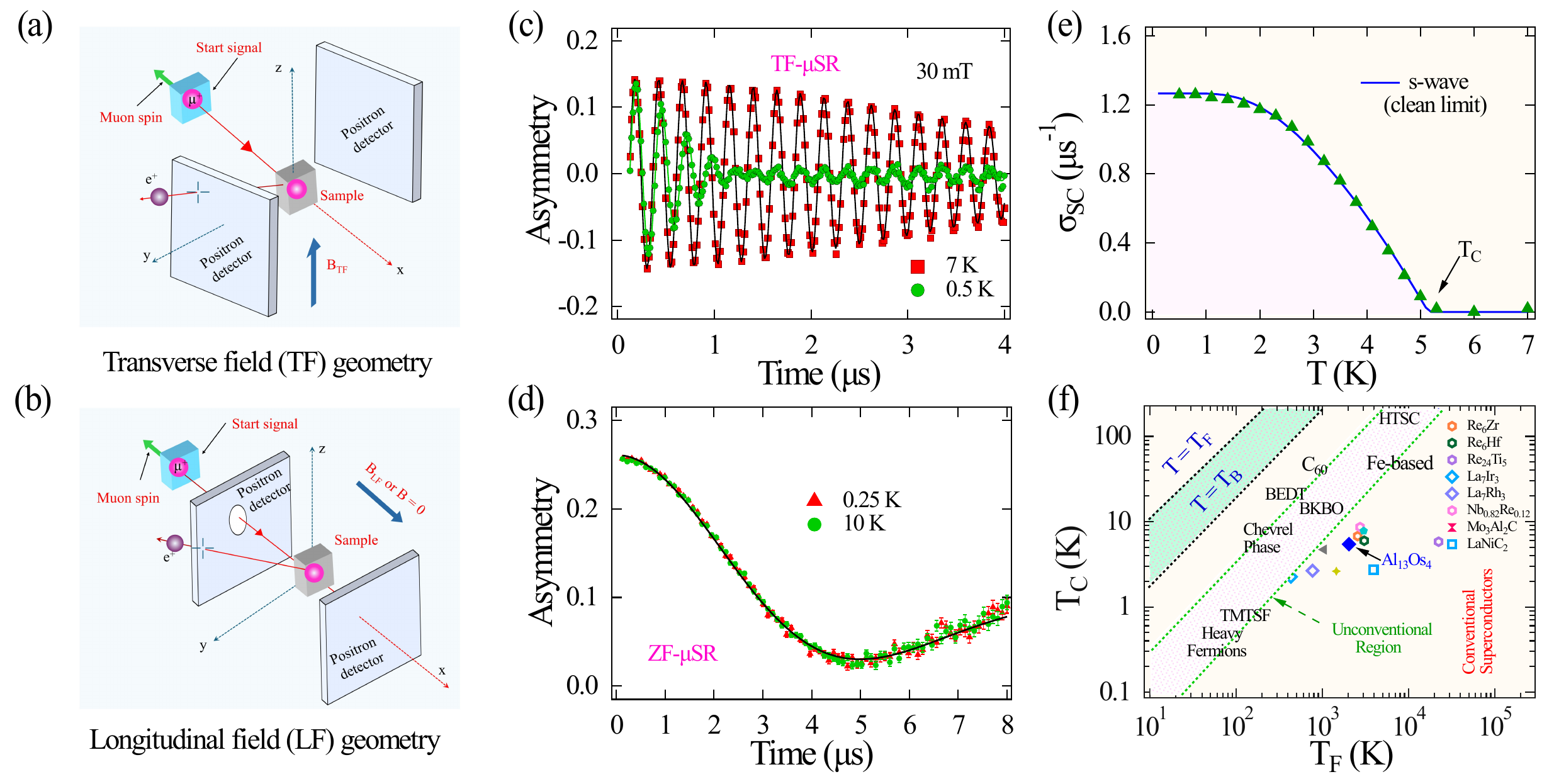}
\caption {\textbf{Microscopic characterization of superconductivity in the decagonal quasicrystal approximant Al$_{13}$Os$_{4}$.} Schematic diagram of the $\mu$SR set up for (a) transverse field (TF) and (b) longitudinal field (LF) configurations. Various directions are marked. (c) TF-$\mu$SR spectra obtained at 30 mT, both above and below superconducting T$_{C}$ with relevant fits. (d) Zero field (ZF)-$\mu$SR spectra above and below T$_{C}$ with a solid line fit representing the product of the static Kubo-Toyabe function and an exponential decay function. (e) Temperature-dependent superconducting relaxation rate ($\sigma_{SC}$) fitted with the isotropic conventional $s$-wave model. (f) Uemura plot between superconducting transition temperature $T_{C}$ and Fermi temperature $T_{F}$ for various types of superconductors.}
\label{Fig2}
\end{figure*}

To further investigate the superconducting gap symmetry and ground-state properties of Al$_{13}$Os$_{4}$, we performed Muon spin rotation and relaxation ($\mu$SR) measurements at ISIS, RAL UK, employing both transverse field (TF) and longitudinal field (LF) geometries, as illustrated in Figs.~\ref{Fig2}(a) and \ref{Fig2}(b).

\textbf{TF-$\mu$SR and fully gapped superconductivity:} TF-$\mu$SR experiments are used to examine the superconducting gap symmetry by measuring the magnetic penetration depth and field distribution in the mixed state. To establish a well-ordered flux line lattice (FLL), experiments were performed under field-cooled conditions. Figure~\ref{Fig2}(c) shows the asymmetry spectra (precession signal) measured above and below $T_{C}$ in an applied field of 30 mT, where the faster relaxation rate at 0.5 K compared 7 K reflects the inhomogeneous field distribution from the FLL. In all temperature ranges, time domain TF-$\mu$SR asymmetry spectra are well described by the Gaussian-damped oscillatory function given as,
\begin{eqnarray}
A(t) = A_{1} exp \left(-\frac{1}{2} \sigma^{2} t^{2}\right) cos(\gamma_{\mu}B_{1}t+\phi)\nonumber \\ + A_{2} cos(\gamma_{\mu}B_{2}t+\phi),
\label{Gaussian oscillatory}
\end{eqnarray}
where $A_{1}$ and $A_{2}$ denote the initial asymmetries corresponding to the sample and non-relaxing background from the silver sample holder, $B_{1}$ and $B_{2}$ are the local magnetic fields sensed by the muons in the sample and sample holder, and $\gamma_{\mu}/2 \pi$ = 135.5 MHz/T is the muon gyromagnetic ratio. $\phi$ denotes the common phase offset. The Gaussian depolarization rate, $\sigma$, is given by $\sigma^{2} = \sigma_{N}^{2} + \sigma_{sc}^{2}$, where $\sigma_{N}$ = 0.219 $\mu s^{-1}$ is the temperature-independent depolarization rate associated with the nuclear dipole moment and $\sigma_{sc}$ represents the depolarization rate from the FLL. The temperature dependence $\sigma_{sc}$ exhibits a plateau at low temperatures. Subsequently, it decreases as the temperature increases before finally reaching zero at $T_{C}$ (Fig.~\ref{Fig2}(e)). To determine the superconducting gap, the temperature-dependent penetration depth $\lambda^{-2}(T)$ $\propto$ $\sigma_{sc}$ was fitted with an isotropic $s$-wave BCS superconductor in the clean limit \cite{brandt2003properties},
\begin{equation}
\frac{\sigma_{sc}(T)}{\sigma_{sc}(0)} = \frac{\lambda^{-2}(T)}{\lambda^{-2}(0)}= 1 + 2 \int_{\Delta(T)}^{\infty} \left(\frac{\partial f}{\partial E} \right) \frac{E dE}{\sqrt{E^{2}-\Delta(T)^{2}}},
\label{s-wave}
\end{equation}
where $\lambda(0)$ is the London penetration depth, $f = [1+ e^{(E/k_{B}T)}]^{-1}$ is the Fermi distribution function, and $\Delta (T)$ is the BCS superconducting gap function given as $\Delta(T) = \Delta_{0}$tanh[1.82$(1.018((T_{C}/T)-1))^{0.51}$]. Fitting of $\lambda^{-2}(T)$ with Eq.~\ref{s-wave} provides a superconducting gap as $\Delta(0)$ = 0.79(3) meV. The normalized gap, $\Delta(0)/k_{B}T_{C}$ = 1.78(6), is consistent to that expected for a BCS weak coupling superconductor. To determine the penetration depth $\lambda^{\mu SR}$(0), the following relation used where $k_{GL} \ge 5$ and $H \le H_{C2}$ \cite{devarakonda2020clean};
\begin{equation}
\sigma_{sc} (\mu s^{-1}) = 4.854 \times 10^{4} (1-h)[1+1.21(1-\sqrt{h})^3] \lambda^{-2},
\label{penetrationdepth}
\end{equation}
with $h = H/H_{C2}(0)$ the reduced field. The calculated value of $\lambda^{\mu}$(0) = 268.6(7) $nm$ closely matches the values from the magnetization as well as those obtained through the analysis of the London penetration depth. Considering the measured residual resistance, we derive a ratio of the Faber-Pippard superconducting coherence length and the electronic transport mean free path ($\xi_{0}$/$l_{e}$) that approaches the clean limit of the superconductor [See SM for individual electronic parameter calculations] \cite{devarakonda2020clean}.

Uemura plot classified superconductors as conventional and unconventional based on the transition temperature $T_C$ and Fermi temperature $T_F$. The $T_{C}$/$T_{F}$ ratio is 0.002 ($T_{C}$ = 5.47(2) K, $T_{F}$ = 2038(9) K; see SM for calculation), marked by a blue square in Fig.~\ref{Fig2}(f) for Al$_{13}$Os$_4$, indicating its location close to the region of unconventional superconductors.\\
\textbf{ZF-$\mu$SR and preserved time-reversal symmetry:} We further employ zero-field (ZF)-$\mu$SR measurements to probe spontaneous magnetization linked to time-reversal symmetry breaking in the superconducting state. The overlapped ZF-$\mu$SR asymmetry spectra below and above $T_{C}$ (Fig.~\ref{Fig2}(d)) show no detectable spontaneous magnetization, confirming the preserved time-reversal symmetry in Al$_{13}$Os$_{4}$. The best description of the ZF spectra is given as,
\begin{equation}
A(t) = A_{0}G_{KT}(t)exp(-\Lambda t) + A_{bg},
\label{KT-ZF1}
\end{equation}
where $A_{0}$ is the initial sample asymmetry and $A_{bg}$ is the non-decaying background from muons stopped in the silver sample holder. The Gaussian static Kubo-Toyabe (KT) function ($G_{KT}(t)$) is given as \cite{hayano1979zero}:
\begin{equation}
G_{KT}(t) = \frac{1}{3}+\frac{2}{3}(1-\Delta^{2}t^{2})exp\left(-\frac{\Delta^{2}t^{2}}{2}\right),
\label{KT-ZF2}
\end{equation}
The relaxation parameter $\Delta$ is associated with randomly oriented, static local fields from nuclear moments, and the electronic relaxation rate, $\Lambda$, both remains nearly constant above and below the superconducting T$_{C}$, indicating preserved time-reversal symmetry in Al$_{13}$Os$_{4}$.

\subsection*{Electronic structure and topology}

\begin{figure*}[ht!]
\includegraphics[width=0.98\linewidth]{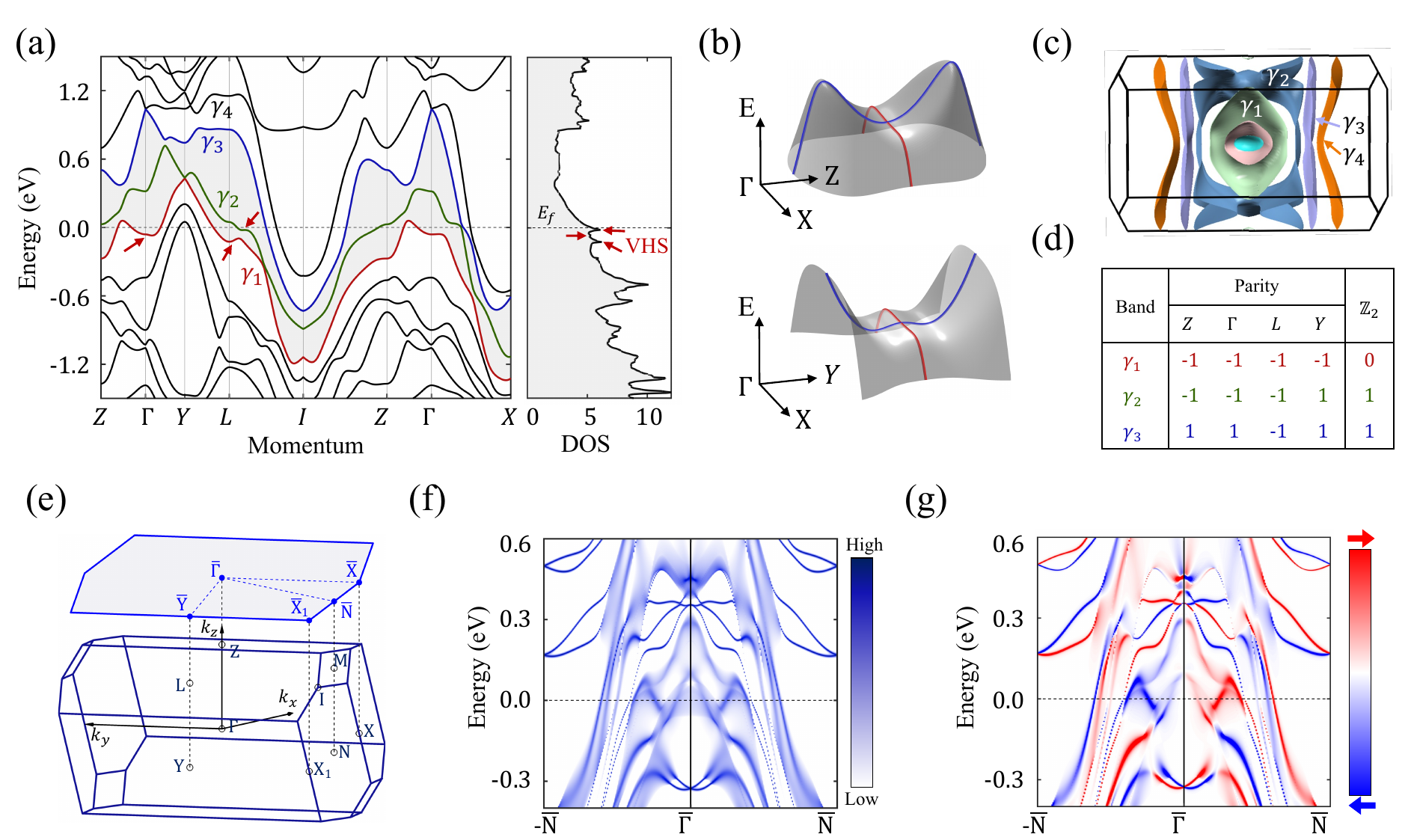}
\caption{\textbf{Nontrivial electronic topology of the decagonal quasicrystal approximant Al$_{13}$Os$_{4}$.} (a) The band structure and density of states (DOS) incorporating spin-orbit coupling. The saddle-points van Hove singularities (vHSs) are indicated by red arrows. $\gamma_1$, $\gamma_2$, $\gamma_3$, and $\gamma_4$ bands crossings the Fermi level are marked. The shaded gray color highlights a continuous band gap between various bands in the Brillouin zone (BZ). (b) Closeup of band structure around $\Gamma$ resolving 3D saddle points with substantial flat band dispersion. (c) Fermi surface with various Fermi sheets in the BZ. (d) Parity and $\mathbb{Z}{_2}$ invariant of $\gamma_1$, $\gamma_2$, and $\gamma_3$ bands. (e) Bulk BZ and (001) surface projected BZ with various high-symmetry points. (f) (001) surface band structure and (g) associated spin-texture along -$\overline{N}-\overline{\Gamma}-\overline{N}$ direction. The spin-momentum-locked nontrivial surface states are observed within the projected bulk bandgap. Red and blue in (g) indicate up and down spin polarizations.}
\label{dft}
\end{figure*}

To examine the electronic topology of decagonal approximant Al$_{13}$Os$_{4}$, we performed first-principles calculations considering experimental parameters with $C2/m$ symmetry. The calculated band structure and density of states (DOS) exhibit metallic behavior, with several bands crossing the Fermi level $E_f$ (Figs.~\ref{dft}(a)-(b)). The four bands, labeled $\gamma_{i=1-4}$, occupy the substantial area of the BZ at the Fermi level. Among these, the $\gamma_{2}$ and $\gamma_{3}$ bands display a 2D character with open Fermi sheets at $E_f$ (Fig.~\ref{dft}(c)). In contrast, the $\gamma_1$ band exhibits 3D saddle-points energy dispersion with a large area flat band close to $E_f$ as shown in Fig.~\ref{dft}(b). At the $\Gamma$ point, the electron (blue) and hole (red) bands cross to form a saddle point with a slowly decaying quadratic dispersion along the $\Gamma-X$ direction. This saddle point gives van Hove singularities (vHSs) near the Fermi level. Such features in the electronic structure are expected to enhance superconducting instability, in agreement with the experimental observations.

Next, we calculate the Sommerfeld-specific heat coefficient and compare it with the experimental data. From the results, we obtain a total DOS at the Fermi level, $D$($E_f$) = 5.38 states eV$^{-1}$ f.u.$^{-1}$. This yields a theoretical Sommerfeld specific-heat coefficient $\gamma_0=\frac{\pi^2 k_B^2}{3} D(E_f)$ of $12.70$ mJ mol$^{-1}$ K$^{-2}$, which is $\sim 60\%$ of the experimentally measured value of $\gamma_n = 21.54(7)$ mJ mol$^{-1}$ K$^{-2}$. This difference infers that electron-mass renormalization due to saddle point could be larger than expected, potentially indicating strong correlation effects. Using the calculated ($\gamma_0$) and experimental ($\gamma_n$) specific-heat coefficients, we estimate the electron-phonon coupling constant ($\lambda_{e-ph}$) using the relation $\frac{\gamma_n}{\gamma_0}=(1+\lambda_{e-ph}$). The calculated $\lambda_{e-ph}$ is 0.69, which is in close agreement with the experimental value of 0.63(1).

Having discussed the electronic properties, we now explore the topological electronic state of Al$_{13}$Os$_{4}$ with spin-orbit coupling (SOC) (see Figs.~\ref{dft}(d)-(g)). The band structure without SOC shows several nodal crossings between different bands near the Fermi level (see SM). These crossings form nodal lines in the bulk BZ. However, due to the lower $C_{2h}$ point-group symmetry and inversion symmetry in Al$_{13}$Os$_{4}$, the inclusion of SOC gaps out these nodal band crossings, resulting in a continuous bandgap at each $k$-point in the BZ. This local bandgap facilitates the calculations of the $\mathcal{Z}_2$ topological invariant similar to insulators \cite{fu2007topological}. In Fig.~\ref{dft}(d), we show the product of parity eigenvalues at various time-reversal invariant points for bands $\gamma_{i=1-3}$~\cite{fu2007topological}. This parity analysis reveals a nontrivial $\mathcal{Z}_2$ invariant when either $\gamma_2$ or $\gamma_3$ bands are considered as valence bands. Figures~\ref{dft}(f)-(g) display the (001) surface band structure and corresponding spin texture. Several spin-momentum-locked surface states are seen to cross the Fermi level. Specifically, within the bulk band-inverted region around $+0.3$ eV, a clear nontrivial Dirac cone surface state appears inside the projected bulk bands. The lower state of this Dirac cone exhibits substantial bandwidth and crosses the Fermi level. The calculated nonzero $\mathcal{Z}_2$ invariant and non-trivial Dirac cone surface states confirm that Al$_{13}$Os$_{4}$ is $\mathcal{Z}_2$ nontrivial. The presence of nontrivial spin-polarized surface states can become superconducting via the bulk proximity effect, leading to unconventional superconductivity in Al$_{13}$Os$_{4}$.

\section*{Summary and discussion}
Our results position Al$_{13}$Os$_{4}$ as a distinctive decagonal QC approximent that exhibits high superconducting $T_C$ and nontrivial electronic topology. It crystallizes in a monoclinic structure with $C2/m$ symmetry group, where two aperiodic layers are stacked along the $\vec{b}$ direction to form a periodic unit cell.  Analysis of resistivity, specific heat, magnetization and $\mu$ SR measurements demonstrates that Al$_{13}$Os$_{4}$ exhibits bulk type-II superconductivity with $T_{C}$ = 5.47(2) K, an upper critical field $H_{C2}$(0) = 1.24(1) T. Zero-field electronic specific heat data align with weak-coupling BCS superconductivity, yielding a superconducting gap of 1.72(1), which closely matches the BCS-type superconductivity gap of 1.78(6) obtained with TF-$\mu$SR measurements. Importantly, the time-reversal symmetry is preserved in the superconducting state within the detection limit of the $\mu$SR. The observed weak-coupling, clean-limit BCS superconductivity with preserved time-reversal symmetry in the superconducting ground state of Al$_{13}$Os$_{4}$ suggests phonon-mediated pairing despite the quasicrystalline atomic arrangement. Our first-principles calculations reveal that Al$_{13}$Os$_{4}$ is a topological metal with a nontrivial $\mathcal{Z}_2$ invariant and vHSs near the Fermi level. 

The aperiodic crystal structure of QCs and their approximants induces both local and global breaks in translational symmetry, which makes a momentum-space perspective inadequate for understanding their properties. The possibility of irregular lattice vibrations arising from aperiodicity may result in unconventional non-BCS superconductivity~\cite{sakai2017superconductivity}. However, the observation of weak-coupling superconductivity with preserved time-reversal symmetry in Al$_{13}$Os$_{4}$ raises important questions about the interplay between aperiodic structures and conventional superconductivity. QCs are also known to suppress superconducting $T_{C}$ when transitioning from approximants to the QC phase. The higher $T_C$ observed in Al$_{13}$Os$_{4}$ may thus attributed to its 2D approximant nature, which retains periodicity in one direction. This contrasts with other 3D QCs such as i-Al-Zn-Mn~\cite{kamiya2018discovery} and the 2D quasicrystalline Ta$_{1.6}$Te~\cite{tokumoto2024superconductivity}, where the superconductivity is more suppressed. 

Moreover, the incorporation of heavy transition metal atoms in Al$_{13}$Os$_{4}$ enhances SOC strength, which plays a crucial role in the realization of non-trivial topological state with nontrivial $\mathcal{Z}_2$, topological surface states and VHS in the energy spectrum. These features are analogous to recently discovered kagome superconductors such as AV$_{3}$Sb$_{5}$ (A = Cs, K)~\cite{ortiz2020cs}. These results strengthen that Al$_{13}$Os$_{4}$ is a promising candidate for exploring topological superconductivity, potentially stabilizing Majorana modes within the vortex cores of its superconducting surface states, which are naturally proximitized by the bulk superconductivity. Note that materials exhibiting both topologically nontrivial surface states and intrinsic superconductivity are rare. Notable example include FeSe$_{1-x}$Te$_{x}$ \cite{zhang2018observation}, doped Bi$_{2}$Se$_{3}$ ~\cite{sasaki2011topological}, and Sn$_{1-x}$In$_{x}$Te~\cite{sasaki2012odd}, all of which show promising signs of unconventional superconductivity. The combination of superconductivity and nontrivial electronic topology in Al$_{13}$Os$_{4}$ presents an exciting platform to explore the interplay between superconductivity, topological effects, and vHSs, and offers new insights into the possible realization of topological superconductivity in QCs and their approximants. 

\section*{Methods}

{\small\textbf{Sample preparation and characterization--} 
Al$_{13}$Os$_{4}$ was synthesized by arc melting high-purity Os ($4N$) and Al ($4N$) in the 13: 4 stoichiometric ratio. The arc melting process was carried out on a water-cooled copper hearth under a high-purity Ar ($4N$) atmosphere. To ensure homogeneity, the alloy was re-melted multiple times after each flip. To examine phase purity and crystal structure characterization, room temperature powder X-ray diffraction was performed using a PANalytical X$^{'}$pert Pro diffractometer using a monochromatic x-ray source with $\lambda$ = 1.5406 $\text{\AA}$ Cu$K_{\alpha}$ radiation. VESTA visualization tool was used to visualize the crystal structure \cite{momma2011vesta}.}\\
{\small\textbf{Superconducting and normal state characterization--}
To investigate the superconducting state, magnetization measurements as a function of temperature and magnetic field were performed using a Quantum Design Superconducting Quantum Interference Device (SQUID) magnetometer in vibrating sample magnetometer (VSM) mode. Electrical transport measurements were conducted using the four-probe technique on a Quantum Design Physical Property Measurement System (PPMS). Heat capacity measurements at zero magnetic field were carried out on the PPMS system using the two-tau relaxation method. A small piece of the sample was affixed to the sapphire holder of the calorimeter using Apiezon-N grease and cooled to 1.9 K. Measurements were then performed in heating mode. To obtain the heat capacity of the sample, the signals from the calorimeter and the grease, measured independently under identical conditions, were subtracted.}\\
{\small\textbf{$\mu$SR experiments--} 
$\mu$SR measurements were performed using the MuSR spectrometer at the ISIS pulsed muon facility, STFC RAL, U.K. $\mu$SR is a microscopic technique for directly probing superconducting materials by measuring spontaneous magnetic fields and vortex state with high sensitivity. It can detect minute variations in internal magnetic fields as small as $\sim$ 1 $\mu$T (detection limit) and magnetic moments down to a few hundredths of a $\mu_{B}$, through the precession of muons with the Larmor frequency $\nu_{\mu} = \gamma_{\mu}/(2\pi)B_{\mu}$ in local magnetic fields at the muon site. The Al$_{13}$Os$_{4}$ system was studied in a zero field (ZF) to detect internal fields. Sample position achieves a near-zero field environment with a tolerance of 1 $\mu$T, which is made possible by three sets of orthogonal coils and an active-field compensation system. Additionally, transverse field (TF) modes, employing a magnetic field perpendicular to the incident muon spin direction, were utilized to explore the gap of a superconductor. For measurement, 3g of powdered sample was mounted on a silver holder with diluted GE varnish and placed in a dilution refrigerator. Nearly 100 \% spin-polarized positive muons were implanted, decaying into positrons and neutrinos after a mean lifetime of 2.2 $\mu$s, and the decayed positrons were collected by four detector groups positioned around the sample position: top, bottom, forward, and backward. The asymmetry of the muon signal is determined using the relation $A(t) = [N_{F}(t)-\alpha N_{B}(t)]/[N_{F}(t)+\alpha N_{B}(t)]$, where $N_{B}(t)$ and $N_{F}(t)$ are the number of detector counts in the forward and backward positions and $\alpha$ is an experiment-specific constant determined from calibration measurements. The reference provides an in-depth explanation of the $\mu$SR technique and detector geometries \cite{hillier2022muon}. The experimental data was analyzed using MANTID software.}\\
{\small\textbf{First-principles calculations--} Electronic structure calculations were performed within the framework of density functional theory based on the projected augmented wave (PAW) method using Vienna \textit{ab-initio} simulation package~\cite{kohn1965self, kresse1999ultrasoft}. The generalized gradient approximation (GGA)~\cite{perdew1996generalized} was used to include the exchange-correlation effects and the spin-orbit coupling was added self-consistently. The kinetic energy cut-off of 400 eV for the plane-wave basis set and a $\Gamma$-centered $11 \times 11 \times 13$ $k$ mesh for Brillouin zone sampling were used. We used experimental lattice constants and relaxed the internal atomic positions until the residual forces on each ion were less than $10^{-2}$ eV\AA$^{-1}$. To explore the topological electronic state, we generated a material-specific tight-binding model Hamiltonian using the VASP2WANNIER90 interface~\cite{mostofi2008wannier90}. We used Os $s$, $d$ and Al $s$, $p$ orbitals to generate the Wannier functions. The surface states and spin texture were obtained using the iterative Green's functions method~\cite{wu2018wanniertools, sancho1985highly}.}

\section*{Data availability}
All data that support the findings of this study are available within the paper and/or Supplemental Materials. Additional data related to this paper may be requested from R.P.S. (rpsingh@iiserb.ac.in).

\bibliographystyle{naturemag}
\bibliography{Main}

\begin{thebibliography}{10}
\expandafter\ifx\csname url\endcsname\relax
  \def\url#1{\texttt{#1}}\fi
\expandafter\ifx\csname urlprefix\endcsname\relax\def\urlprefix{URL }\fi
\providecommand{\bibinfo}[2]{#2}
\providecommand{\eprint}[2][]{\url{#2}}

\bibitem{shechtman1984metallic}
\bibinfo{author}{Shechtman, D.}, \bibinfo{author}{Blech, I.},
  \bibinfo{author}{Gratias, D.} \& \bibinfo{author}{Cahn, J.~W.}
\newblock \bibinfo{title}{Metallic phase with long-range orientational order
  and no translational symmetry}.
\newblock \emph{\bibinfo{journal}{Phys. Rev. Lett.}}
  \textbf{\bibinfo{volume}{53}}, \bibinfo{pages}{1951} (\bibinfo{year}{1984}).

\bibitem{deguchi2012quantum}
\bibinfo{author}{Deguchi, K.} \emph{et~al.}
\newblock \bibinfo{title}{Quantum critical state in a magnetic quasicrystal}.
\newblock \emph{\bibinfo{journal}{Nat. Mater.}} \textbf{\bibinfo{volume}{11}},
  \bibinfo{pages}{1013} (\bibinfo{year}{2012}).

\bibitem{hafner1992electronic}
\bibinfo{author}{Hafner, J.} \& \bibinfo{author}{Kraj{\v{c}}{\'\i}, M.}
\newblock \bibinfo{title}{Electronic structure and stability of quasicrystals:
  Quasiperiodic dispersion relations and pseudogaps}.
\newblock \emph{\bibinfo{journal}{Phys. Rev. Lett.}}
  \textbf{\bibinfo{volume}{68}}, \bibinfo{pages}{2321} (\bibinfo{year}{1992}).

\bibitem{nayak2012bulk}
\bibinfo{author}{Nayak, J.} \emph{et~al.}
\newblock \bibinfo{title}{Bulk electronic structure of quasicrystals}.
\newblock \emph{\bibinfo{journal}{Phys. Rev. Lett.}}
  \textbf{\bibinfo{volume}{109}}, \bibinfo{pages}{216403}
  (\bibinfo{year}{2012}).

\bibitem{tamura2021experimental}
\bibinfo{author}{Tamura, R.} \emph{et~al.}
\newblock \bibinfo{title}{Experimental observation of long-range magnetic order
  in icosahedral quasicrystals}.
\newblock \emph{\bibinfo{journal}{J. Am. Chem. Soc.}}
  \textbf{\bibinfo{volume}{143}}, \bibinfo{pages}{19938}
  (\bibinfo{year}{2021}).

\bibitem{chen2020higher}
\bibinfo{author}{Chen, R.}, \bibinfo{author}{Chen, C.-Z.},
  \bibinfo{author}{Gao, J.-H.}, \bibinfo{author}{Zhou, B.} \&
  \bibinfo{author}{Xu, D.-H.}
\newblock \bibinfo{title}{Higher-order topological insulators in
  quasicrystals}.
\newblock \emph{\bibinfo{journal}{Phys. Rev. Lett.}}
  \textbf{\bibinfo{volume}{124}}, \bibinfo{pages}{036803}
  (\bibinfo{year}{2020}).

\bibitem{fan2022topological}
\bibinfo{author}{Fan, J.} \& \bibinfo{author}{Huang, H.}
\newblock \bibinfo{title}{Topological states in quasicrystals}.
\newblock \emph{\bibinfo{journal}{Front. Phys.}} \textbf{\bibinfo{volume}{17}},
  \bibinfo{pages}{13203} (\bibinfo{year}{2022}).

\bibitem{ghadimi2021topological}
\bibinfo{author}{Ghadimi, R.}, \bibinfo{author}{Sugimoto, T.},
  \bibinfo{author}{Tanaka, K.} \& \bibinfo{author}{Tohyama, T.}
\newblock \bibinfo{title}{Topological superconductivity in quasicrystals}.
\newblock \emph{\bibinfo{journal}{Phys. Rev. B}}
  \textbf{\bibinfo{volume}{104}}, \bibinfo{pages}{144511}
  (\bibinfo{year}{2021}).

\bibitem{sakai2019exotic}
\bibinfo{author}{Sakai, S.} \& \bibinfo{author}{Arita, R.}
\newblock \bibinfo{title}{Exotic pairing state in quasicrystalline
  superconductors under a magnetic field}.
\newblock \emph{\bibinfo{journal}{Phys. Rev. Research}}
  \textbf{\bibinfo{volume}{1}}, \bibinfo{pages}{022002} (\bibinfo{year}{2019}).

\bibitem{araujo2019conventional}
\bibinfo{author}{Ara{\'u}jo, R.~N.} \& \bibinfo{author}{Andrade, E.~C.}
\newblock \bibinfo{title}{Conventional superconductivity in quasicrystals}.
\newblock \emph{\bibinfo{journal}{Phys. Rev. B}}
  \textbf{\bibinfo{volume}{100}}, \bibinfo{pages}{014510}
  (\bibinfo{year}{2019}).

\bibitem{liu2024nematic}
\bibinfo{author}{Liu, Y.-B.}, \bibinfo{author}{Zhou, J.} \&
  \bibinfo{author}{Yang, F.}
\newblock \bibinfo{title}{Nematic superconductivity and its critical vestigial
  phases in the quasi-crystal}.
\newblock \emph{\bibinfo{journal}{Phys. Rev. Lett.}}
  \textbf{\bibinfo{volume}{133}}, \bibinfo{pages}{136002}
  (\bibinfo{year}{2024}).

\bibitem{sakai2017superconductivity}
\bibinfo{author}{Sakai, S.}, \bibinfo{author}{Takemori, N.},
  \bibinfo{author}{Koga, A.} \& \bibinfo{author}{Arita, R.}
\newblock \bibinfo{title}{Superconductivity on a quasiperiodic lattice:
  Extended-to-localized crossover of cooper pairs}.
\newblock \emph{\bibinfo{journal}{Phys. Rev. B}} \textbf{\bibinfo{volume}{95}},
  \bibinfo{pages}{024509} (\bibinfo{year}{2017}).

\bibitem{graebner1987specific}
\bibinfo{author}{Graebner, J.~E.} \& \bibinfo{author}{Chen, H.~S.}
\newblock \bibinfo{title}{Specific heat of an icosahedral superconductor,
  \text{Mg$_{3}$Zn$_{3}$Al$_{2}$}}.
\newblock \emph{\bibinfo{journal}{Phys. Rev. Lett.}}
  \textbf{\bibinfo{volume}{58}}, \bibinfo{pages}{1945} (\bibinfo{year}{1987}).

\bibitem{wagner1988specific}
\bibinfo{author}{Wagner, J.~L.}, \bibinfo{author}{Biggs, B.~D.},
  \bibinfo{author}{Wong, K.~M.} \& \bibinfo{author}{Poon, S.~J.}
\newblock \bibinfo{title}{Specific-heat and transport properties of alloys
  exhibiting quasicrystalline and crystalline order}.
\newblock \emph{\bibinfo{journal}{Phys. Rev. B}} \textbf{\bibinfo{volume}{38}},
  \bibinfo{pages}{7436} (\bibinfo{year}{1988}).

\bibitem{kamiya2018discovery}
\bibinfo{author}{Kamiya, K.} \emph{et~al.}
\newblock \bibinfo{title}{Discovery of superconductivity in quasicrystal}.
\newblock \emph{\bibinfo{journal}{Nat. Commun.}} \textbf{\bibinfo{volume}{9}},
  \bibinfo{pages}{154} (\bibinfo{year}{2018}).

\bibitem{tokumoto2024superconductivity}
\bibinfo{author}{Tokumoto, Y.} \emph{et~al.}
\newblock \bibinfo{title}{Superconductivity in a van der waals layered
  quasicrystal}.
\newblock \emph{\bibinfo{journal}{Nat. Commun.}} \textbf{\bibinfo{volume}{15}},
  \bibinfo{pages}{1529} (\bibinfo{year}{2024}).

\bibitem{terashima2024anomalous}
\bibinfo{author}{Terashima, T.} \emph{et~al.}
\newblock \bibinfo{title}{Anomalous upper critical field in the quasicrystal
  superconductor \text{Ta$_{1.6}$Te}}.
\newblock \emph{\bibinfo{journal}{npj Quantum Materials}}
  \textbf{\bibinfo{volume}{9}}, \bibinfo{pages}{56} (\bibinfo{year}{2024}).

\bibitem{Kraus_TQC1}
\bibinfo{author}{Kraus, Y.~E.}, \bibinfo{author}{Lahini, Y.},
  \bibinfo{author}{Ringel, Z.}, \bibinfo{author}{Verbin, M.} \&
  \bibinfo{author}{Zilberberg, O.}
\newblock \bibinfo{title}{Topological states and adiabatic pumping in
  quasicrystals}.
\newblock \emph{\bibinfo{journal}{Phys. Rev. Lett.}}
  \textbf{\bibinfo{volume}{109}}, \bibinfo{pages}{106402}
  (\bibinfo{year}{2012}).

\bibitem{Kraus_TQC2}
\bibinfo{author}{Kraus, Y.~E.} \& \bibinfo{author}{Zilberberg, O.}
\newblock \bibinfo{title}{Topological equivalence between the fibonacci
  quasicrystal and the harper model}.
\newblock \emph{\bibinfo{journal}{Phys. Rev. Lett.}}
  \textbf{\bibinfo{volume}{109}}, \bibinfo{pages}{116404}
  (\bibinfo{year}{2012}).

\bibitem{cain2020layer}
\bibinfo{author}{Cain, J.~D.}, \bibinfo{author}{Azizi, A.},
  \bibinfo{author}{Conrad, M.}, \bibinfo{author}{Griffin, S.~M.} \&
  \bibinfo{author}{Zettl, A.}
\newblock \bibinfo{title}{Layer-dependent topological phase in a
  two-dimensional quasicrystal and approximant}.
\newblock \emph{\bibinfo{journal}{Proc. Natl. Acad. Sci. USA}}
  \textbf{\bibinfo{volume}{117}}, \bibinfo{pages}{26135}
  (\bibinfo{year}{2020}).

\bibitem{singh2023topology}
\bibinfo{author}{Singh, B.}, \bibinfo{author}{Lin, H.} \&
  \bibinfo{author}{Bansil, A.}
\newblock \bibinfo{title}{Topology and symmetry in quantum materials}.
\newblock \emph{\bibinfo{journal}{Advanced Materials}}
  \textbf{\bibinfo{volume}{35}}, \bibinfo{pages}{2201058}
  (\bibinfo{year}{2023}).

\bibitem{zhang1995al13}
\bibinfo{author}{Zhang, B.}, \bibinfo{author}{Gramlich, V.} \&
  \bibinfo{author}{Steurer, W.}
\newblock \bibinfo{title}{\text{Al$_{13-x}$(Co$_{1- y}$Ni$_{y}$)$_{4}$}, a new
  approximant of the decagonal quasicrystal in the \text{Al-Co-Ni} system}.
\newblock \emph{\bibinfo{journal}{Z. Kristallogr}}
  \textbf{\bibinfo{volume}{210}}, \bibinfo{pages}{498} (\bibinfo{year}{1995}).

\bibitem{li1995structure}
\bibinfo{author}{Li, X.~Z.}, \bibinfo{author}{Shi, N.~C.}, \bibinfo{author}{Ma,
  Z.~S.}, \bibinfo{author}{Ma, X.~L.} \& \bibinfo{author}{Kuo, K.~H.}
\newblock \bibinfo{title}{Structure of \text{Al$_{11}$Co$_{4}$}, a new
  monoclinic approximant of the \text{Al-Co} decagonal quasicrystal}.
\newblock \emph{\bibinfo{journal}{Philos. Mag. Lett.}}
  \textbf{\bibinfo{volume}{72}}, \bibinfo{pages}{79} (\bibinfo{year}{1995}).

\bibitem{steurer1993structure}
\bibinfo{author}{Steurer, W.}, \bibinfo{author}{Haibach, T.},
  \bibinfo{author}{Zhang, B.}, \bibinfo{author}{Kek, S.} \&
  \bibinfo{author}{L{\"u}ck, R.}
\newblock \bibinfo{title}{The structure of decagonal
  \text{Al$_{70}$Ni$_{15}$Co$_{15}$}}.
\newblock \emph{\bibinfo{journal}{Acta Cryst.}} \textbf{\bibinfo{volume}{49}},
  \bibinfo{pages}{661} (\bibinfo{year}{1993}).

\bibitem{brandt2003properties}
\bibinfo{author}{Brandt, E.~H.}
\newblock \bibinfo{title}{Properties of the ideal ginzburg-landau vortex
  lattice}.
\newblock \emph{\bibinfo{journal}{Phys. Rev. B}} \textbf{\bibinfo{volume}{68}},
  \bibinfo{pages}{054506} (\bibinfo{year}{2003}).

\bibitem{devarakonda2020clean}
\bibinfo{author}{Devarakonda, A.} \emph{et~al.}
\newblock \bibinfo{title}{Clean \text{2D} superconductivity in a bulk van der
  waals superlattice}.
\newblock \emph{\bibinfo{journal}{Science}} \textbf{\bibinfo{volume}{370}},
  \bibinfo{pages}{231} (\bibinfo{year}{2020}).

\bibitem{hayano1979zero}
\bibinfo{author}{Hayano, R.~S.} \emph{et~al.}
\newblock \bibinfo{title}{Zero-and low-field spin relaxation studied by
  positive muons}.
\newblock \emph{\bibinfo{journal}{Phys. Rev. B}} \textbf{\bibinfo{volume}{20}},
  \bibinfo{pages}{850} (\bibinfo{year}{1979}).

\bibitem{fu2007topological}
\bibinfo{author}{Fu, L.} \& \bibinfo{author}{Kane, C.~L.}
\newblock \bibinfo{title}{Topological insulators with inversion symmetry}.
\newblock \emph{\bibinfo{journal}{Phys. Rev. B}} \textbf{\bibinfo{volume}{76}},
  \bibinfo{pages}{045302} (\bibinfo{year}{2007}).

\bibitem{ortiz2020cs}
\bibinfo{author}{Ortiz, B.~R.} \emph{et~al.}
\newblock \bibinfo{title}{\text{CsV$_{3}$Sb$_{5}$}: A \text{Z$_{2}$}
  topological kagome metal with a superconducting ground state}.
\newblock \emph{\bibinfo{journal}{Phys. Rev. Lett.}}
  \textbf{\bibinfo{volume}{125}}, \bibinfo{pages}{247002}
  (\bibinfo{year}{2020}).

\bibitem{zhang2018observation}
\bibinfo{author}{Zhang, P.} \emph{et~al.}
\newblock \bibinfo{title}{Observation of topological superconductivity on the
  surface of an iron-based superconductor}.
\newblock \emph{\bibinfo{journal}{Science}} \textbf{\bibinfo{volume}{360}},
  \bibinfo{pages}{182} (\bibinfo{year}{2018}).

\bibitem{sasaki2011topological}
\bibinfo{author}{Sasaki, S.} \emph{et~al.}
\newblock \bibinfo{title}{Topological superconductivity in
  \text{Cu$_{x}$Bi$_{2}$Se$_{3}$}}.
\newblock \emph{\bibinfo{journal}{Phys. Rev. Lett.}}
  \textbf{\bibinfo{volume}{107}}, \bibinfo{pages}{217001}
  (\bibinfo{year}{2011}).

\bibitem{sasaki2012odd}
\bibinfo{author}{Sasaki, S.} \emph{et~al.}
\newblock \bibinfo{title}{Odd-parity pairing and topological superconductivity
  in a strongly spin-orbit coupled semiconductor}.
\newblock \emph{\bibinfo{journal}{Phys. Rev. Lett.}}
  \textbf{\bibinfo{volume}{109}}, \bibinfo{pages}{217004}
  (\bibinfo{year}{2012}).

\bibitem{momma2011vesta}
\bibinfo{author}{Momma, K.} \& \bibinfo{author}{Izumi, F.}
\newblock \bibinfo{title}{\text{VESTA} 3 for three-dimensional visualization of
  crystal, volumetric and morphology data}.
\newblock \emph{\bibinfo{journal}{J. Appl. Cryst.}}
  \textbf{\bibinfo{volume}{44}}, \bibinfo{pages}{1272} (\bibinfo{year}{2011}).

\bibitem{hillier2022muon}
\bibinfo{author}{Hillier, A.~D.} \emph{et~al.}
\newblock \bibinfo{title}{Muon spin spectroscopy}.
\newblock \emph{\bibinfo{journal}{Nat. Rev. Methods Primers}}
  \textbf{\bibinfo{volume}{2}}, \bibinfo{pages}{4} (\bibinfo{year}{2022}).

\bibitem{kohn1965self}
\bibinfo{author}{Kohn, W.} \& \bibinfo{author}{Sham, L.~J.}
\newblock \bibinfo{title}{Self-consistent equations including exchange and
  correlation effects}.
\newblock \emph{\bibinfo{journal}{Phys. Rev.}} \textbf{\bibinfo{volume}{140}},
  \bibinfo{pages}{A1133} (\bibinfo{year}{1965}).

\bibitem{kresse1999ultrasoft}
\bibinfo{author}{Kresse, G.} \& \bibinfo{author}{Joubert, D.}
\newblock \bibinfo{title}{From ultrasoft pseudopotentials to the projector
  augmented-wave method}.
\newblock \emph{\bibinfo{journal}{Phys. Rev. B}} \textbf{\bibinfo{volume}{59}},
  \bibinfo{pages}{1758} (\bibinfo{year}{1999}).

\bibitem{perdew1996generalized}
\bibinfo{author}{Perdew, J.~P.}, \bibinfo{author}{Burke, K.} \&
  \bibinfo{author}{Ernzerhof, M.}
\newblock \bibinfo{title}{Generalized gradient approximation made simple}.
\newblock \emph{\bibinfo{journal}{Phys. Rev. Lett.}}
  \textbf{\bibinfo{volume}{77}}, \bibinfo{pages}{3865} (\bibinfo{year}{1996}).

\bibitem{mostofi2008wannier90}
\bibinfo{author}{Mostofi, A.~A.} \emph{et~al.}
\newblock \bibinfo{title}{wannier90: A tool for obtaining maximally-localised
  wannier functions}.
\newblock \emph{\bibinfo{journal}{Comput. Phys. Commun.}}
  \textbf{\bibinfo{volume}{178}}, \bibinfo{pages}{685} (\bibinfo{year}{2008}).

\bibitem{wu2018wanniertools}
\bibinfo{author}{Wu, Q.}, \bibinfo{author}{Zhang, S.}, \bibinfo{author}{Song,
  H.-F.}, \bibinfo{author}{Troyer, M.} \& \bibinfo{author}{Soluyanov, A.~A.}
\newblock \bibinfo{title}{Wanniertools: An open-source software package for
  novel topological materials}.
\newblock \emph{\bibinfo{journal}{Comput. Phys. Commun.}}
  \textbf{\bibinfo{volume}{224}}, \bibinfo{pages}{405} (\bibinfo{year}{2018}).

\bibitem{sancho1985highly}
\bibinfo{author}{Sancho, M. P.~L.}, \bibinfo{author}{Sancho, J. M.~L.},
  \bibinfo{author}{Sancho, J.~L.} \& \bibinfo{author}{Rubio, J.}
\newblock \bibinfo{title}{Highly convergent schemes for the calculation of bulk
  and surface green functions}.
\newblock \emph{\bibinfo{journal}{J. Phys. F: Met. Phys.}}
  \textbf{\bibinfo{volume}{15}}, \bibinfo{pages}{851} (\bibinfo{year}{1985}).

\end{thebibliography}

\section*{Acknowledgments}

P.K.M. acknowledges the funding agency Council of Scientific and Industrial Research (CSIR), Government of India, for providing the SRF fellowship (Award No: 09/1020(0174)/2019-EMR-I). R.P.S. acknowledges the Science and Engineering Research Board, Government of India, for the Core Research Grant No. CRG/2023/000817 and ISIS, STFC, UK, for providing beamtime for the $\mu$SR experiments. The work at TIFR Mumbai is supported by the Department of Atomic Energy of the Government of India under Project No. 12-R\&D-TFR-5.10- 0100 and benefited from the HPC resources of TIFR Mumbai. 

\section*{Author contributions}
RPS conceived the project. PKM, A, SJ, RKK, and RPS prepared the samples and conducted the characterization. RPS, ADH, RS, and PKM carried out the $\mu$SR experiments and data analysis. RV and BS performed the first-principles calculations and symmetry analysis. PKM, A, RV, BS, and RPS wrote the manuscript, with input from all authors. This work is supervised by RPS (experiment) and BS (theory).

\section*{Competing interests} The authors declare no competing interests.

\newpage

\newpage

\newpage

%\section*{Additional information}

\end{document}